\documentstyle[aps,prb,eqsecnum,psfig,floats,preprint]{revtex}
\begin{document}
%@twocolumnfalse\endcsname
\tightenlines
\draft

\title
{Strongly nonlinear convection in binary fluids:\\
Minimal model for extended states using symmetry decomposed modes
\footnote{Dedicated to Prof. W. G{\"o}tze on the occasion of his
sixtieth birthday}
}

\author{St.~Hollinger and M.~L\"ucke}
\address{Institut f\"ur Theoretische Physik, Universit\"at des
Saarlandes, Postfach 151150, D 66041 Saarbr\"ucken, Germany}
%\date{1 November 1996}
\maketitle
 
\begin{abstract}
Spatially extended stationary and traveling states in the strongly nonlinear
regime of 
convection in layers of binary fluid mixtures heated from below are described
by a few--mode--model. It is derived from the proper hydrodynamic balance
equations including experimentally relevant boundary conditions with a
non--standard Galerkin approximation that uses numerically obtained, 
symmetry decomposed modes. Properties of the model are elucidated and compared
with full numerical solutions of the field equations.
\end{abstract}
 

% 47.20.-k : Hydrodynamic stability
% 47.10.+g : General theory (of fluid dynamics)
% 03.40.Gc : Classical mechanics of continuous media:
%            general mathematical aspects
%
%\maketitle

%\vskip2pc]
\narrowtext

%****************************************************************
%*                                                              *
%*      begin of text                                           *
%*                                                              *
%****************************************************************

%\advance\baselineskip by 6 pt
%\advance\hsize by 92mm

\newcommand{\Nabla}{\mbox{\bf\boldmath $\nabla$}}

%\input 1.tex
%****************************************************************
%*                                                              *
%*      INTRODUCTION						*
%*                                                              *
%****************************************************************

\section{Introduction}
Convection in binary miscible fluids like, e. g.,
ethanol--water is a paradigmatic system for studying instabilities,
bifurcations, complex
spatiotemporal behaviour, and turbulence. Furthermore, it is
sufficiently simple to allow for controlled experiments
and the governing field equations are well known. So
recently a lot of research activities [1--13] have been
directed towards investigating the enormous variety of
pattern forming behaviour in this system. The richness
of spatio temporal phenomena in binary fluid mixtures 
stems from a feed--back loop between the fields of 
velocity, concentration, and temperature. Let us start
with the velocity field: The convective flow is driven
by the buoyancy force field which itself is determined
by variations of the temperature and of the concentration
field. The latter are on the one hand generated via the
thermodiffusive Soret effect by temperature gradients and
on the other hand they are reduced by concentration 
diffusion and by mixing through the convective flow.
Since these changes influence the buoyancy which drives
the flow the feed back loop is closed. 

In this article we concentrate on spatially extended 
convection states of straight parallel rolls that occur
either as a horizontally traveling wave (TW) or in the
form of stationary "overturning" convection (SOC) rolls. 
In particular, we derive a model which is able to describe
also stable TWs in the strongly nonlinear regime and which goes 
beyond earlier approximations that are applicable only
to the weakly nonlinear regime. We present and explore
here the properties of a minimal Galerkin model that is
based on a symmetry decomposition of numerically obtained
fields. It reproduces the typical bifurcation scenario 
obtained from simulations and experiments.

For fluid parameters typically realized in mixtures of water and about 10 wt.\%
ethanol, an oscillatory, subcritical onset of convection is observed. It is
connected by an unstable TW branch with a saddle node bifurcation giving rise
to stable, strongly nonlinear TW states. At a certain
Rayleigh number, the phase velocity of these waves vanishes and the SOC
branch of stable stationary states can be observed. Along the TW bifurcation
branch which is shown in detail in Figs.~1 and 4 the concentration changes its
structure from lateral homogeneity and
vertically linear layering in the basic state over plateau--like
distributions in fast TWs to boundary layer dominated, slowly traveling
waves. The contrast between two adjacent TW rolls
is strongly related to the phase velocity of the TW and it vanishes with this
velocity. So, SOCs do not show such a concentration contrast.
In SOCs adjacent rolls are mirror images of each other and
they are separated from another and from the top and
bottom plate only by thin boundary layers. The
latter are a typical phenomenon for
convection of weakly diffusing scalars. To capture this behaviour of
the concentration field our 
model contains more degrees
of freedom of the concentration field than of the velocity and of
the temperature field. 

Our article is organized as follows: The {\bf second} section presents the
system, the fields needed for its description, their governing equations
with the explanation of the relevant fluid and control
parameters. The {\bf third} section shows which symmetries in TW and SOC convection
occur and how they change at the bifurcations. Furthermore,
a discussion of the symmetry decomposed balance equations and of
their numerically found solutions is given. Out of this data we extract
in the {\bf fourth} section a basis for a non--standard Galerkin approximation.
Its results are presented in bifurcation diagrams of different order parameters
in comparison with the correct solutions of the system. We also discuss
predictions of our model concerning the dependence of the bifurcation
topology on the strength of the Soret coupling.

%\input 2.tex
%****************************************************************
%*                                                              *
%*      System		                                        *
%*                                                              *
%****************************************************************

\section{System}
The fields needed for the description of straight parallel
rolls in a horizontal fluid layer exposed to a vertical gravitational
acceleration $g$ are temperature $T$, concentration $C$, and velocity
${\rm\bf u}=(u,0,w)$. Here, $u$ is the velocity field in the lateral
$x$--direction and $w$ is the velocity in the vertical $z$--direction.
Fluid parameters are the Prandtl
number $\sigma=\nu/\kappa$, the Lewis number $L=D/\kappa$, and the separation
ratio $\psi=-\beta k_{T}/(\alpha T_{0})$. $\nu$ is the kinematic
viscosity, $\kappa$ the thermal diffusivity, $D$ the diffusion coefficient,
$\alpha=-\big(1/\rho\big)\,\big(\partial\rho/\partial T\big)$ and
$\beta=-\big(1/\rho\big)\,\big(\partial\rho/\partial C\big)$
thermal and solutal expansion
coefficients, $\rho$ the fluid's density, $k_T$ the thermodiffusion ratio,
and $T_0$ ($C_0$) the mean temperature (concentration) in the system. We
scale lengths by the height $d$ of the fluid layer,
times by the vertical
thermal diffusion time $d^2/\kappa$, temperatures by the imposed vertical
temperature difference $\Delta T$, and concentrations by $\Delta T\alpha/\beta$.
Our control parameter is the Rayleigh number 
$R=\alpha g d^3 \Delta T /(\nu\kappa)$. Mostly we use the reduced
Rayleigh number
$r=R/R_c^0$ where $R_c^0$ denotes the critical Rayleigh number for the onset
of convection in a pure fluid ($R_c^0 = 1707.762$; $R_c^0\simeq 1747.28$
in our model).

We start with the hydrodynamic field equations in a nondimensionsal
version using the scales defined above. They represent balances for
momentum, heat, and concentration
\begin{mathletters}
\begin{eqnarray}
\partial_t{\rm\bf u} + {\rm\bf u}\cdot\Nabla{\rm\bf u} & = &
 - \Nabla p
 + \sigma\left\{R\left[\left(T-T_0\right)+\left(C-C_0\right)\right]
   {\rm\bf e}_z
 + \nabla^2{\rm\bf u}\right\}\\
\partial_t T + {\rm\bf u}\cdot\Nabla T & = & \nabla^2 T\\
\partial_t C + {\rm\bf u}\cdot\Nabla C & = & L \nabla^2 C - L \psi \nabla^2 T.
\end{eqnarray}
\end{mathletters}
Assuming an incompressible fluid --- this is reasonable as long as the
convective velocities in the fluid are small in comparison with the
velocity of the sound --- the density is constant and the continuity 
equation as the balance equation for the total mass leads to
$$\Nabla\cdot{\rm\bf u} = 0\ \ \ \ .$$
In the equations (2.1), the temporal changes of a field are given by
convective transport of that field on the left hand side of the equation.
Apart from the advective rate of change the momentum (2.1a) changes by
pressure gradients ($-\Nabla p$) and dissipative effects
($\sigma\nabla^2{\rm\bf u}$). The gravitational acceleration acts as an
external force ($R\left[\left(T-T_0\right)+\left(C-C_0\right)\right]{\rm\bf e}_z$) on
the fluid. At that point, density fluctuations due to temperature and
concentration fluctuations may not be ignored since they represent the
driving of the system.

The heat balance  (2.1b) takes advective (${\rm\bf u}\cdot\Nabla T$) and
diffusive ($\nabla^2T$) transport
into account. In the case of the concentration
balance (2.1c), the cross coupling of the temperature to the concentration
($-L\psi\nabla^2T$) is a relevant effect. It is called thermodiffusion or
Soret effect [1,2]. For negative separation ratios $\psi$, it acts in such
a way that temperature gradients induce antiparallel concentration
gradients; positive values of $\psi$ lead to parallel ones. If the
Soret coupling $\psi$ vanishes, as it occurs in ethanol water mixtures of a
concentration of 0 and 28 wt.\%, the mixture behaves just like a pure
fluid.

From these remarks, the stability behaviour may be inferred: For
negative separation ratios, the unstable layering of the density is
reduced, for positive ones it is increased compared with the situation in a
pure fluid. Thus, mixtures with $\psi<0$ ($\psi>0$) have
larger (smaller) critical Rayleigh numbers than the pure fluid.
Below these threshold values, a motionless (${\rm\bf u}\equiv 0$) state
with the laterally homogeneous conductive profiles
\begin{eqnarray*}
T_{\rm cond}(z) & = & T_0 - z\\
C_{\rm cond}(z) & = & C_0 - \psi z
\end{eqnarray*}
of temperature and concentration, respectively, is a stable solution of
eqs. (2.1). Above threshold this solution becomes unstable to convection.
It is convenient to use 
\begin{mathletters}
\begin{eqnarray}
\theta  & = & T(x,z;t)-T_{cond}(z)\\
 c & = & C(x,z;t)-C_{cond}(z)\ \ \ \ .
\end{eqnarray}
\end{mathletters}
as convective deviations of temperature and concentration from their conductive 
profiles for straight rolls in $y$-direction.

In this paper we discuss convective roll solutions of (2.1) which have a lateral
wave number of  $k=\pi$ being close to the critical values around $3.116$. 
These solutions are either stationary or they represent traveling waves. The 
latter are stationary in a frame comoving with the phase velocity $v$ of
the traveling wave relative to the laboratory frame. In the comoving frame
($v\equiv0$ for states being stationary in the laboratory frame),
we introduce the streamfunction $\Phi(x,z)$ which describes the
incompressible twodimensional convective flow
\begin{equation}
{\rm\bf u}(x,z) = ( u(x,z) , 0 , w(x,z) ) = 
                  (-\partial_z,0,\partial_x)\Phi(x,z)
\end{equation}
in the $x-z$ plane perpendicular to the roll axes. The velocity field 
${\rm\bf u}(x,z)$ (2.3) guarantees incompressibility,
$\Nabla\cdot{\rm\bf u} = 0$, by construction.  

The traveling and stationary states are solutions of the time independent
field equations
\begin{mathletters}
\begin{eqnarray}
{\cal J}(\Phi,\nabla^2\Phi)
 & = & \sigma\left[R\partial_x\left(\theta+c\right) + \nabla^4\Phi\right]\\
{\cal J}(\Phi,\theta)
 & = & \partial_x\Phi + \nabla^2\theta\\
{\cal J}(\Phi,c)
 & = & \psi\partial_x\Phi + L\nabla^2\left(c-\psi\theta\right)
\end{eqnarray}
\end{mathletters}
in the comoving frame. They can directly be deduced from (2.1). Herein,
we used the Jacobian
$$ {\cal J}(\Phi,f) := \big(\partial_x\Phi\big)\partial_zf
                     - \big(\partial_z\Phi\big)\partial_xf
                  \equiv \big({\rm\bf u}\cdot\Nabla\big) f $$
to describe the convective nonlinearity.

The boundary conditions at $z=\pm1/2$ in the comoving frame are for all $x$
\begin{eqnarray}
\makebox[4cm][l]{\rm no\ slip} u & =\ -\partial_z\Phi & =\ - v\ \ \ ,\nonumber\\
\makebox[4cm][l]{} w & =\ \ \partial_x\Phi & =\ 0\ \ \ ,\nonumber\\
\makebox[4cm][l]{\rm perfectly heat conducting} & \theta & =\ 0 \ \ \ ,\ {\rm and}\\
\makebox[4cm][l]{\rm impermeable} & \partial_z(c-\psi\theta) & = \ 0 \ \ \ .\nonumber
\end{eqnarray}

Numerical simulations [13] and a full scale Galerkin analysis with many 
trigonometric modes for the fields [15] have shown that
the contribution from
lateral Fourier modes higher than the first one in the velocity field is
only in a range of 1\% of the first one and that these higher velocity modes
are unnecessary for reproducing the
bifurcation topology and the peculiar structure of the concentration field,
respectively. The analyses [13,15] have also proved the irrelevance of
a lateral meanflow in the lateral velocity field. Therefore, we
start our approximations with the ansatz
\begin{equation}
\Phi(x,z) = {w_{\rm max}\over k\,{\cal C}_1(0)}\sin kx \, {\cal C}_1(z) + vz
          =: \tilde{\Phi}(x,z) + vz 
\end{equation}
for the streamfunction. Here, ${\cal C}_1(z)$ is the first even
Chandrasekhar function [16], $w_{\rm max}$ the maximal
flow amplitude, and $v$ the phase velocity of the traveling state. These
two velocities are the only degrees of freedom of the velocity field. By
the above ansatz for the streamfunction we fix the origin
$x=0, z=0$ of the cooordinate system to a point of maximal upflow
velocity.

%\input 3.tex
%****************************************************************
%*                                                              *
%*      SYMMETRIES						*
%*                                                              *
%****************************************************************

\section{Symmetries}
In this section we discuss symmetries of the fields in the coordinate
system defined in the last section and we perform a symmetry
decomposition of the fields and their balance equations. After this,
we analyse the relative weight of contributions from
different symmetries in the solutions of the field equations.

\subsection{Decomposition}
Nonlinear states of stationary convection
(SOCs) have definite parity [13] under the operation $x\rightarrow -x$:
positive parity for the fields $w$, $\theta$, and $c$, and a negative one for 
$\Phi$ and $u$. Only the linear
critical fields at the stationary bifurcation threshold
have in addition definite vertical
parity under the operation $z\rightarrow -z$: positive
parity for the fields $w$, $\Phi$, $\theta$, and $c$, and a negative for $u$.

Traveling wave states with non vanishing phase velocity have no
definite lateral parity. Only that TW just on the merging point $r^*$ of the
branches of SOCs and TWs has, of course, the same symmetry as the
stationary states. At the point of the Hopf bifurcation out of the
basic state, the linear
critical fields have definite vertical parity, which is
destroyed by the action of the convective nonlinearity in every nonlinear
state. 

The temperature and concentration fields are decomposed in four parts each,
corresponding to the four different combinations of vertical and lateral
parity
\begin{mathletters}
\begin{eqnarray}
\theta  & = & \theta^{++} + \theta^{+-} + \theta^{-+} + \theta^{--} \\
c & = & c^{++} + c^{+-} + c^{-+} + c^{--}\ \ \ \ . 
\end{eqnarray}
\end{mathletters}
Here, the first (second) superscript denotes the parity in lateral (vertical)
direction. The symmetries themselves will be labelled by
${\cal S}^{\pm\pm}$ in the same way. A symmetry decomposition of the velocity field or of the
streamfunction is not necessary, since their spatial variation is fixed
by (2.4). The decomposition of the balance equation for the concentration
(2.3c) yields the following four equations after inserting the streamfunction
$\tilde{\Phi}$ (2.4) into (2.3c)
\begin{mathletters}
\begin{eqnarray}
{\cal J}(\tilde{\Phi},c^{+-})-v\,\partial_x c^{-+}  & = &
 L \nabla^2(c^{++}-\psi\theta^{++})\ +\ \psi\,\partial_x\tilde{\Phi}\\
{\cal J}(\tilde{\Phi},c^{++})-v\,\partial_x c^{--} & = &
 L \nabla^2(c^{+-}-\psi\theta^{+-}) \\
{\cal J}(\tilde{\Phi},c^{--})-v\,\partial_x c^{++} & = &
 L \nabla^2(c^{-+}-\psi\theta^{-+})\\
{\cal J}(\tilde{\Phi},c^{-+})-v\,\partial_x c^{+-} & = &
 L \nabla^2(c^{--}-\psi\theta^{--})\ \ \ \ .
\end{eqnarray}
\end{mathletters}%
The decomposition of the heat balance equation
(2.3b) can be performed in a completely
analogous way. Using the ansatz (2.4) for the streamfunction the momentum
balance can be split into two parts: one for the symmetry ${\cal S}^{++}$
and one for ${\cal S}^{-+}$
\begin{mathletters}
\begin{eqnarray}
 - v\,\partial_x\nabla^2\tilde{\Phi} & =  &
 R\,\sigma\,\partial_x\left(\theta^{-+}+c^{-+}\right)\\
0 & = & R\,\partial_x\left(\theta^{++}+c^{++}\right) +
  \nabla^4\tilde{\Phi}\ \ \ \ .
\end{eqnarray}
\end{mathletters}
All together, eight partial differential equations must be solved for the
temperature and concentration fields. The two degrees of freedom of the
velocity field, $w_{\rm max}$ and $v$, are determined by algebraic
equations which result from projecting the equations (3.3) onto the
fixed spatial modes of the streamfunction (2.4).

\subsection{Numerical solution}
The above equations have been solved by a multi--mode Galerkin method
[15] that implies a full representation of the fields. In particular, it
resolves the peculiar spatial structure in the concentration
field [12,13] in a sufficient way.
%\begfig 1
Fig.~1 shows the resulting bifurcation diagrams of TWs (full lines) and of
SOCs (dashed lines). The TW solution branch bifurcates at the Hopf
bifurcation threshold $r_{\rm osc}$ with the Hopf frequency $\omega_{\rm H}$
out of the quiescent heat conducting state and it merges at $r^*$ with the SOC
branch.
%\begfig 2
For these TWs we show in Fig.~2 how the different symmetry decomposed
contributions to the TW concentration field $c(x,z)$ vary as one moves
along the TW bifurcation branch $w_{\rm max}=0$, i.~e.~from the bifurcation
out of the conductive state, to the end of the TW states at $r^*$. 
We measure the portion of a symmetry $c^{\pm\pm}(x,z)$ in $c(x,z)$ by 
${\langle c^{\pm\pm} | c \rangle}/{\langle c^2 \rangle}$
using global averaging as the scalar product. 
Thus, $c(x,z)$ contains at $r_{\rm osc}$ ($w_{\rm max}=0$)
only contributions of the symmetries ${\cal S}^{-+}$ (short dashed line)
and ${\cal S}^{++}$ (solid line). This reflects the fact that the critical
eigenfunctions have positive vertical parity. However,
already in weakly nonlinear
states the portion of the symmetry ${\cal S}^{+-}$ exceeds
that of ${\cal S}^{++}$ as a result of
the strong increase of the zeroth lateral Fouriermode by the action of
the convective nonlinearity. Near to the point where the TW phase velocity
and the maximal
convective velocity are equal and which has been identified
[17] as the transition from weakly to strongly nonlinear
TW convection the two symmetries ${\cal S}^{+-}$ and
${\cal S}^{-+}$ have the same
portion in $c(x,z)$. Together, they make up 99.9\% of it. Finally,
at the SOC--TW--transition where the TW transforms into an
SOC state there are no more contributions
from ${\cal S}^{-+}$ and ${\cal S}^{--}$.

A correct stability analysis of the conductive state needs test functions
of the symmetries ${\cal S}^{-+}$ and
${\cal S}^{++}$ whereas the determination of the
SOC--TW--transition which can be looked upon as a lateral parity breaking
bifurcation of SOCs containing ${\cal S}^{++}$ and ${\cal S}^{+-}$
requires ${\cal S}^{-+}$ and ${\cal S}^{--}$.
Therefore, in principle all symmetries are necessary to describe
the two transitions in the system
in an adequate manner. The yet unmentioned symmetry ${\cal S}^{--}$ 
(long dashed line in Fig.~2) contributes at
most 0.06\% of $c(x,z)$ and is clearly the smallest portion in $c(x,z)$ 
all over the bifurcation diagram. That is the reason why we think
that modes of this symmetry --- their influence will be discussed 
in Fig.~4 --- are more or less irrelevant for
the complex bifurcation scenario in the convection of binary fluid
mixtures.

%\input 4.tex
%****************************************************************
%*                                                              *
%*      NON-STANDARD-GALERKIN-APPROXIMATION			*
%*                                                              *
%****************************************************************

\section{Non--standard Galerkin approximation}
\subsection{Mode selection}
The simplest approach is to consider only one mode per symmetry class and
to discard the ${\cal S}^{--}$ part in the fields.
This is the essence of older
standard Galerkin approximations [14]. They yield
linear relations between the square of the TW--frequency $\omega$,
the square of the convection amplitude $w_{\rm max}$ and the reduced 
Rayleigh number $r=R/R_c^0$
\begin{mathletters}
\begin{eqnarray}
\omega^2 & =&  \alpha(r-r_{\rm osc})\ +\ \omega_{\rm H}^2\\
w_{\rm max}^2 &  = & \beta(r-r_{\rm osc})\ \ \ \ .
\end{eqnarray}
\end{mathletters}%
Here, $\omega_{\rm H}$ is the Hopf frequency at the
oscillatory stability threshold $r_{\rm osc}$
of the basic state. These linear relations
reproduce the onset of oscillatory convection and its weakly nonlinear
properties in an adequate manner. However,
they are not able to describe a saddle
node bifurcation and stable, strongly nonlinear TW convection.
In fact, it was shown [18] that all Galerkin approximations using
only one mode for each of the three symmetries ${\cal S}^{++}$,
${\cal S}^{-+}$, and ${\cal S}^{+-}$
yield linear relations like (4.1) --- independent of the spatial shape of the 
modes.

Because of the smallness of the portion of ${\cal S}^{--}$ in $c(x,z)$ the
simplest extension is to take
{\it two\/} modes for each of the three relevant
symmetries into account. This works reasonably well,
as we will show below. Thus, such a
model has to be looked upon as minimal model that is able to describe
linear, weakly nonlinear, and strongly nonlinear states in binary
fluid convection.

In order to provide adequate modes we use
numerical solutions of the system as a basis for our
non--standard Galerkin approximation. As one is interested in the
description of an upper stable TW branch and the saddle node bifurcation
we symmetry decompose two states of the upper
(numerically determined) TW branch and then orthonormalize them by a
Gram--Schmidt--method using global averaging as scalar product.
%$$\langle\,f(x,z)\,g(x,z)\,\rangle = {k\over 2\pi}\int_{-\pi/k}^{\pi/k}\!\!dx
%\int_{-1/2}^{1/2}\!\!dz\ f^*(x,z) g(x,z)\ \ \ \ .$$

%\begfig 3
For fluid parameters that are typically realized in mixtures of ethanol
and water ($L=0.01$, $\sigma=10$, $\psi=-0.25$) we use a fast TW near the saddle
($\omega=4$) and a slow one near the SOC--TW--transition ($\omega=0.125$).
Both frequencies are small compared with the Hopf frequency of about
11.3. The resulting basis is displayed in Fig.~3. Since the structure of the
temperature field changes only slightly along the 
whole bifurcation branch it is
sufficient to consider only one temperature mode per symmetry. The three
most important
contributions to $\theta$ are, like in the concentration field,
the symmetries ${\cal S}^{++}$, ${\cal S}^{-+}$, and ${\cal S}^{+-}$.
Their contributions
are very well described by the modes of the classical Lorenz model
[19] and those of related approximations [14].

\subsection{Nonlinear stable TW states}
Using the basis shown in Fig.~3 and the balance equations (2.3)
one can derive
a Galerkin model for convection in binary mixtures.
This model has six or eight degrees of freedom for
the description of the concentration field depending on
whether the contribution of the symmetry
${\cal S}^{--}$ is contained or not. The temperature field is represented
by three or four modes. As usual, projecting the field equations,
e.~g.~(3.2), on the spatial modes of the respective fields yields a system
of algebraic equations for the mode amplitudes.
The order parameters of the velocity field,
$w_{\rm max}$ and $\omega=vk$, are computed from two algebraic 
equations for the velocity field that result after projection from (3.3).
Here, we only want to discuss the bifurcation diagrams of the TWs, since
those of the SOCs are well enough reproduced by earlier approximations
except for quantitative values of $M$ and their stability properties.

%\begfig 4
Fig.~4 shows from top to bottom the TW bifurcation diagrams of
Nusselt number
$$ N\ =\ \langle ({\rm\bf u}T-\Nabla T)\!\cdot\!{\rm\bf e}_z
	 \rangle_x\big|_{z=-1/2} 
    \ =\ 1 -\langle \partial_z\theta\rangle_x\big|_{z=-1/2}\ \ , $$
TW frequency $\omega$, and the reduced concentration variance   
$$ M\ =\ \sqrt{{\langle C^2\rangle \over \langle C_{cond}^2
                   \rangle}}\
  =\ {2\sqrt{3} \over |\psi |} \sqrt{\langle C^2\rangle}\ \ \ \ .$$
These order parameters have been evaluated for extended models
including ${\cal S}^{--}$ modes (dotted lines) and neglecting them
(dashed lines) in comparison
with the "true" solution of the system (solid lines). The first thing 
to notice is that for a very good reproduction of the stable TW branch (the upper one in
$N$, the lower ones in $M$ and $\omega$) the symmetry ${\cal S}^{--}$
may not be neglected. Nevertheless, ${\cal S}^{--}$
is not necessary for a qualitative correct
bifurcation scenario of a Hopf bifurcation, a saddle node bifurcation and a
SOC--TW--transition on the stable, strongly nonlinear branch. This was not
the case in earlier Galerkin models [14]
connecting the basic state with the unstable SOC--branch by a TW--branch.

The positions of the SOC--TW--transitions given by the model including the
${\cal S}^{--}$ symmetry and neglecting it differ as can be seen in Fig.~4.
This can be explained as follows. This transition may be interpreted
as a stability threshold of SOCs whose
lateral positive parity gets broken there. Since
SOCs contain both ${\cal S}^{++}$ and ${\cal S}^{+-}$
they need to be tested for stability against
disturbances of both ${\cal S}^{-+}$ and ${\cal S}^{--}$.
If one neglects ${\cal S}^{--}$ the stability
analysis is incomplete leading to different values for the threshold.

The correct bifurcation of the concentration variance $M$ is due to an
adequate description of the concentration field $C(x,z)$ which is
nearly $0$ in SOCs all over the convection cell except for small boundary
layers between two adjacent rolls and at top and bottom plate so that the
variance around its mean value is very small compared to the 
large concentration gradients in the basic state.

\subsection{Unstable TW solutions}
The unstable TW branches resulting from the models
differ from the exact ones [17], however without
destroying  the subcritical Hopf bifurcation topology --- compare the dashed
and
dotted lower unstable branches of $N$  up to the saddle in Fig.~4a
with the full line (for $\omega$ and $M$ the upper branch corresponds
to the unstable solution and is not shown in Fig.~4b and 4c).
The reason for these
differences is that the basic state is not tested for stability
by the correct critical eigenfunctions. 
Since the oscillatory stability problem in binary fluid mixtures
cannot be formulated as a variational priniciple as it is the case
for the stationary one in a pure fluid [16] there is no need that an
approximate stability threshold is higher than the exact one.

As mentioned already the
position of the Hopf bifurcation does not depend on the presence of the
symmetry ${\cal S}^{--}$
since the basic state needs only to be tested for stability
against disturbances of the symmetries ${\cal S}^{++}$ and
${\cal S}^{-+}$. However, the initial slope of,
e.~g., the Nusselt number $N$ being a nonlinear property
involves modes of the symmetry ${\cal S}^{--}$ that are driven
by the critical modes with the symmetries
${\cal S}^{++}$ and ${\cal S}^{-+}$ via the convective nonlinearity.
All this can be
seen in Fig.~4a where the dotted and dashed lines
have a common bifurcation threshold out of the
basic state ($N=1$) but spread in the nonlinear regime ($N>1$). 
A qualitatively better reproduction of the unstable TW branch is obtained by
changing the basis of the Galerkin model, namely by using states of the
correct unstable TW branch. Then, the approximation quality of the
stable states which are the main topic of this work deteriorates. 

\subsection{Dependence on the separation ratio $\psi$}
%\begfig 5
Numerical simulations [13] have shown that the concentration
distribution depends only slightly on the separation ratio $\psi$ for
the small Lewis numbers $L=O(0.01)$ and large Prandtl numbers $\sigma=O(10)$
that are typically realized in mixtures
of ethanol and water. This insensitiveness
of the concentration field towards changes in $\psi$ suggests that
our above presented models also describe convection in binary mixtures
with values of $\psi$ different to those where the basis is 
extracted. 

The changes in the
bifurcation topology caused by varying the strength $\psi$ of the Soret
coupling may be described by tracing out
the $\psi$--dependences of the SOC and
TW saddle node bifurcations $r_{\rm SOC}^s$ and
$r_{\rm TW}^s$, and of the SOC--TW--transition $r^*$. This is shown in
Fig.~5 for the model containing all symmetries (lines) in
comparison with the full representation of the fields (symbols). 
Both saddles of the SOC and TW bifurcation diagrams procured by
our non--standard model are in 
good quantitative agreement with the results of the
full representation. The model only deviates for weak Soret couplings
$\psi>-0.04$. This can be explained by the smoother
variations of the concentration field for those couplings (see
e.~g.~[13]). The same holds for the SOC--TW--transition $r^*$ which
in addition differs for strong Soret couplings $\psi<-0.3$. This is due
to the fact that the SOC--TW--transition can be interpreted as a boundary
layer instability [20] which therefore is very sensitive to
the thickness of the boundary layers in the concentration field.
However, the thickness of the boundary layer is fixed
in a model with fixed modes. Nevertheless, the model
is able to describe the $\psi$--dependence of both TW-- and
SOC--bifurcation scenario quite well.

%\input 5.tex
%****************************************************************
%*                                                              *
%*      CONCLUSION						*
%*                                                              *
%****************************************************************

\section{Conclusion}
In this article we have presented a model describing traveling wave
convection in binary fluid mixtures. Their complex bifurcation scenario is
reproduced by a Galerkin method using two numerically
obtained exact solutions of the system as a
basis. These solutions for fixed fluid parameters are symmetry
decomposed in order to meet the symmetry properties of the bifurcations and
of the states in the system. We have shown that there exists a minimal number of
degrees of freedom for the description of the concentration field. Our
model uses this minimal number and reproduces numerically and
experimentally observed states in binary fluid mixtures very well. 
In addition, it describes the dependence of the bifurcation topology on the
strength of the Soret coupling quite well.
Furthermore, it shows that in  the
temperature and velocity field the modes of the classical Lorenz model
[19] are sufficient. Finally, the model gives useful hints for a
derivation of an approximation using analytical test
functions instead of the here applied numerical ones.

\acknowledgments
This work was supported by Deutsche For\-schungs\-ge\-mein\-schaft. 
Discussions with W.~Barten, P.~B\"uchel, and H.~W.~M\"ul\-ler are
gratefully acknowledged as well as a graduate scholarship of the
Saarland for one of us (SH).

\begin{figure}[hbt]
\centerline{\psfig{figure=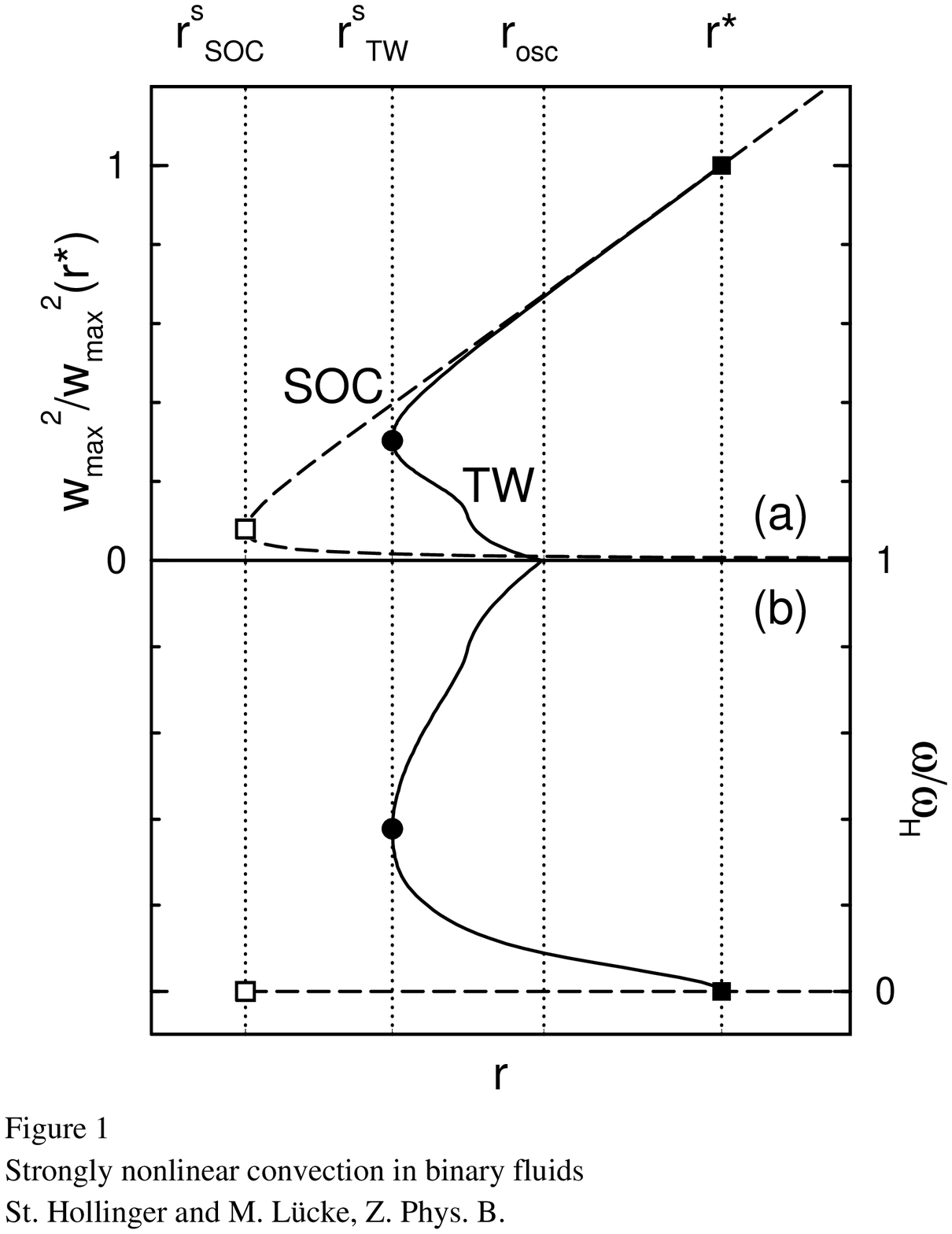,width=150mm}}
\caption{Bifurcation diagrams of reduced convection flow intensity (a)
           and frequency (b) in TW (solid lines) and SOC
           (dashed lines) solutions in a binary mixture with $\psi=-0.25$,
           $L=0.01$ and $\sigma=10$.}
\label{fig:1}
\end{figure}
\begin{figure}
\centerline{\psfig{figure=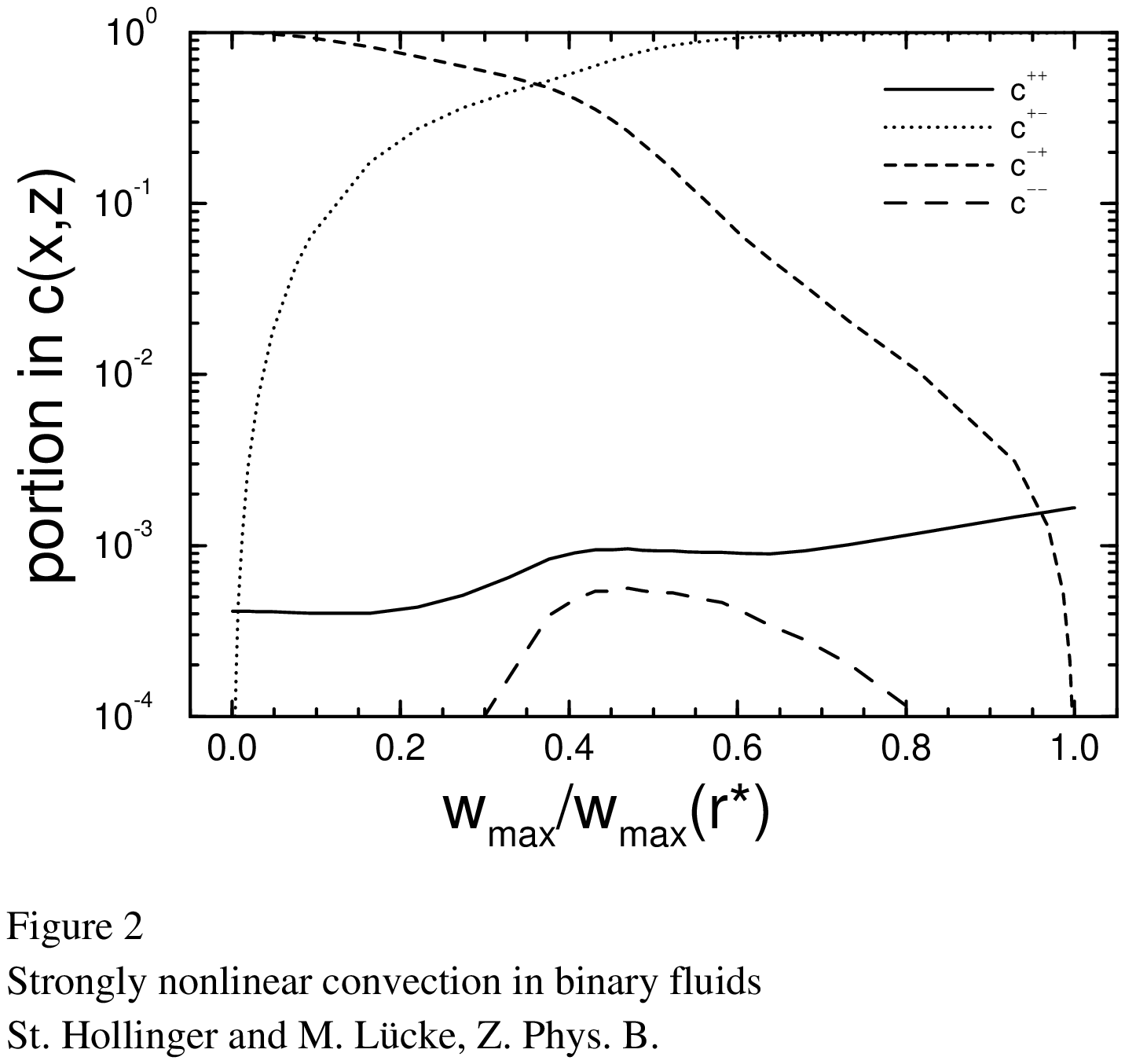,width=180mm}}
\caption{The deviations $c(x,z)$ of the concentration field from of its
           conductive profile are decomposed into four symmetry classes
           having definite parity under the operations $x\rightarrow -x$ and
           $z\rightarrow -z$. The relative weight of the respective components
           $c^{\pm\pm}$ in the complete field $c(x,z)$ are plotted as a
           function of the maximal flow amplitude $w_{\rm max}$ between the
           Hopf bifurcation ($w_{\rm max}=0$) and the SOC--TW--transition
           ($w_{\rm max}=w_{\rm max}(r^*)$). Parameters are $L=0.01$,
           $\sigma=10$, and $\psi=-0.25$.}
\label{fig:2}
\end{figure}
\begin{figure}
\centerline{\psfig{figure=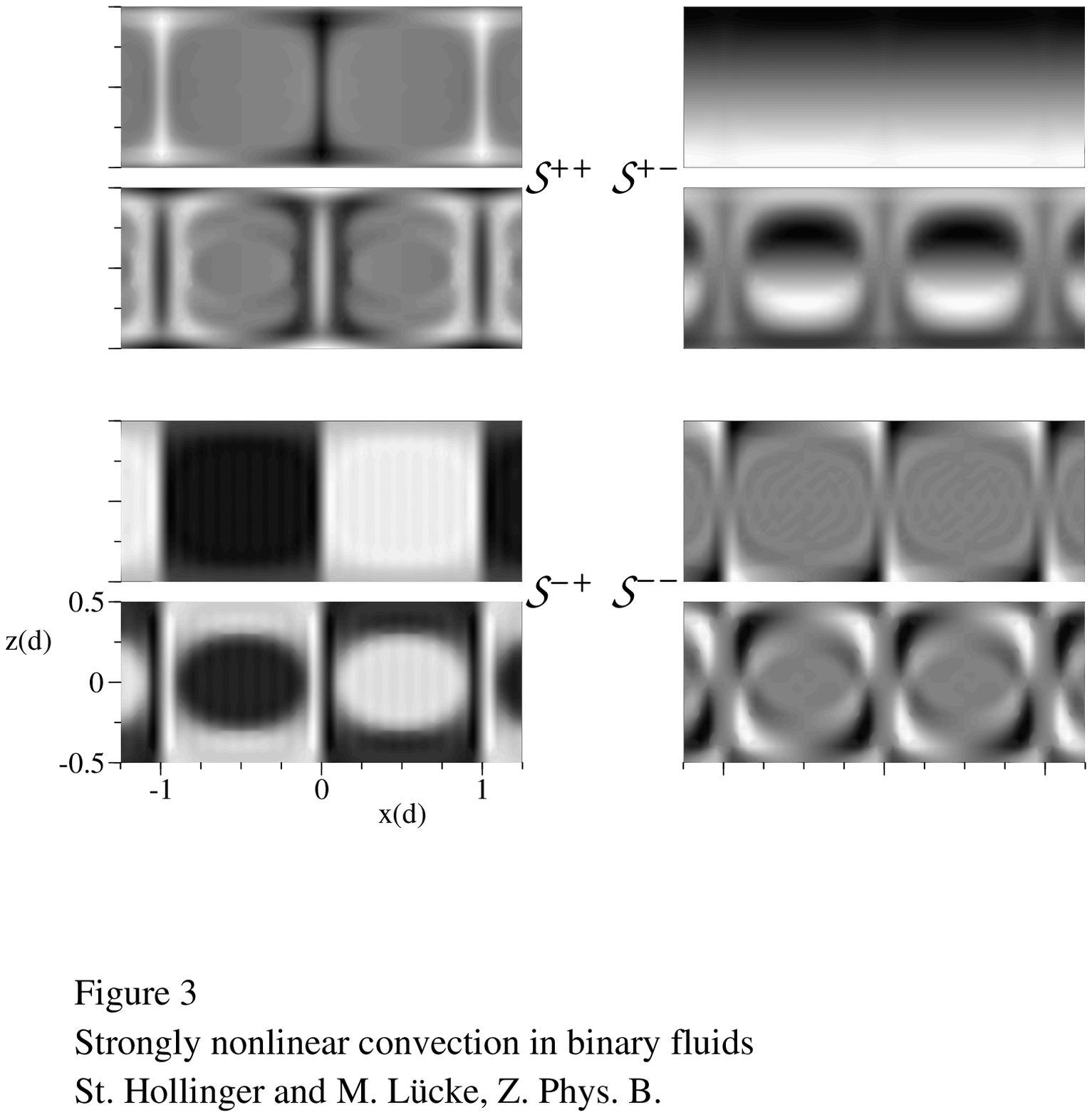,width=140mm}}
\caption{Orthonormalized basis for the description of the concentration
           field in the $x$--$z$--plane. The wave number is $k=\pi$ so that
           the lateral peridicity length is 2. The four different
           symmetries are denoted by ${\cal S}^{\pm\pm}$ at the side of the
           four blocks containing two basis elements each. The upper ones of
           each block are real states of the system ($\omega=0.125$)
           whereas the lower ones are linear combinations of two states
           resulting from the orthonormalization procedure.}
\label{fig:3}
\end{figure}
\begin{figure}
\centerline{\psfig{figure=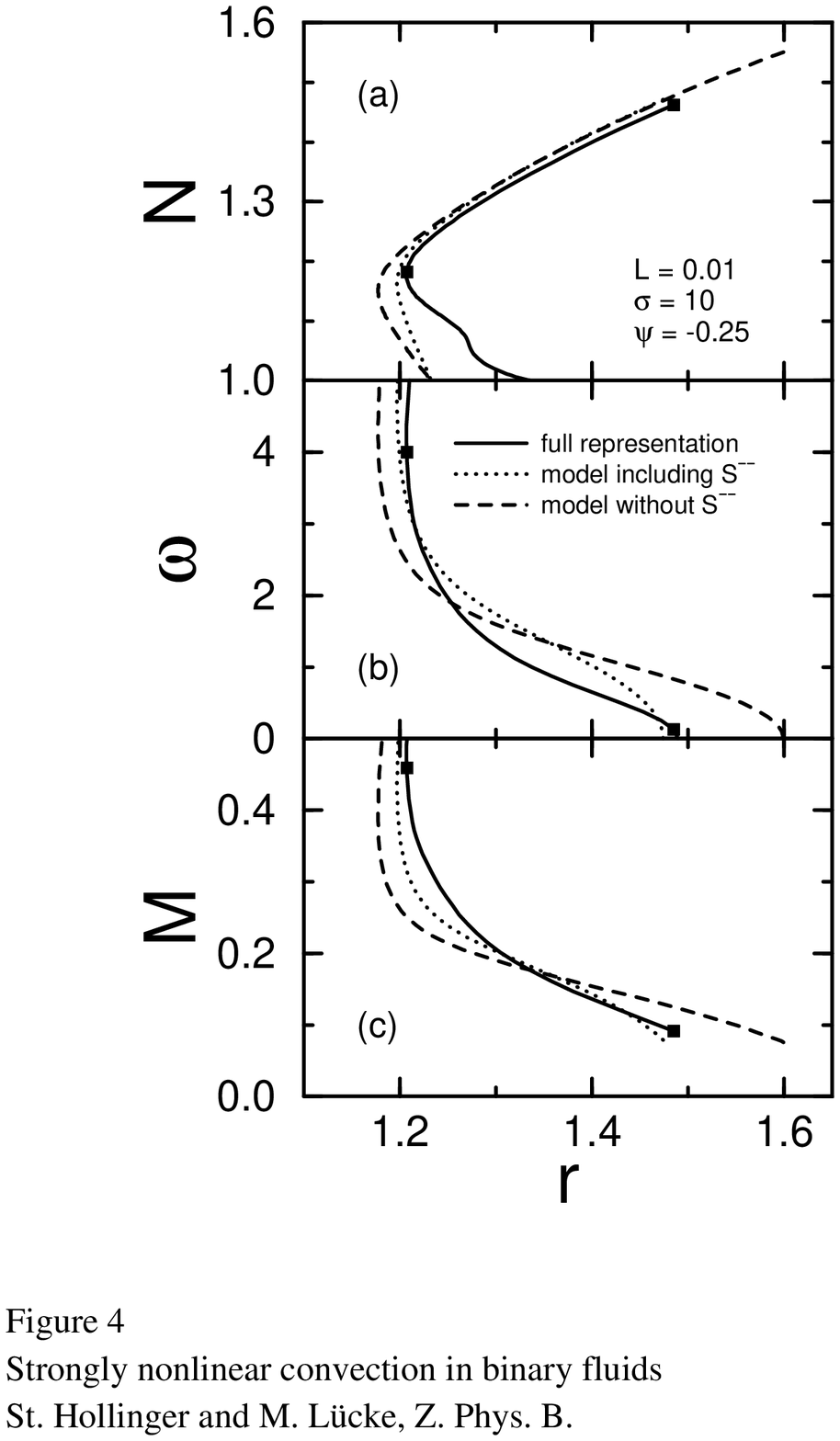,width=150mm}}
\caption{Bifurcation diagrams of Nusselt number $N$ (a), TW frequency
           $\omega$ (b) and concentration variance $M$ (c) in the framework
           of Galerkin models with numerical basis
           including the symmetry ${\cal S}^{--}$ (dotted line) or
           discarding it
           (dashed line) in comparison with results of a many mode analysis
           (solid line). The filled squares show the two states which where
           used to construct a basis for the models. Parameters are
           $L=0.01$, $\sigma=10$, $\psi=-0.25$, and $k=\pi$.}
\label{fig:4}
\end{figure}
\begin{figure}
\centerline{\psfig{figure=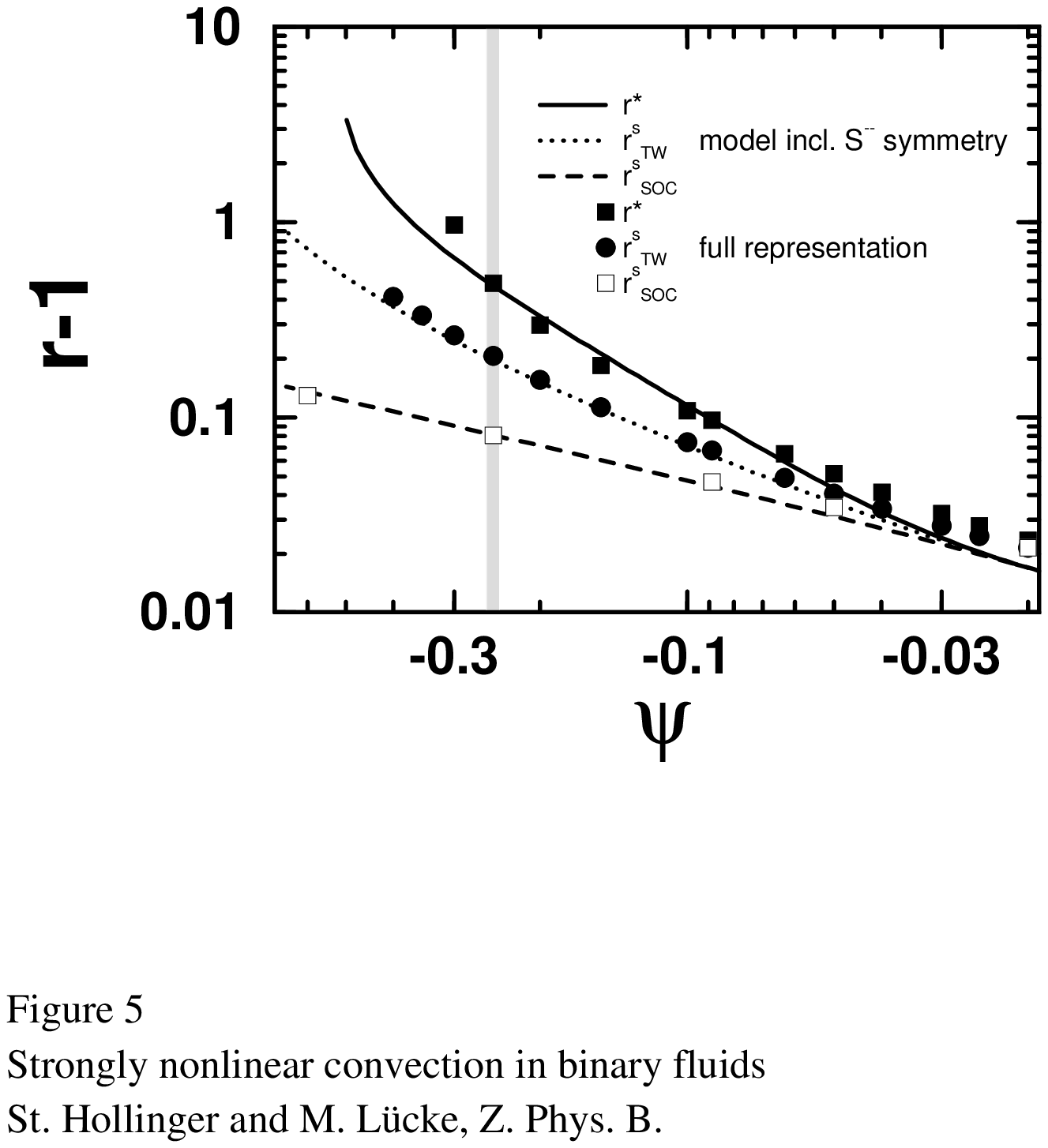,width=150mm}}
\caption{Phase diagrams of the here presented
           non--standard Ga\-ler\-kin model (lines) in comparison with the
           results of a full representation (symbols). The basis of the
           model was fixed at $\psi=-0.25$ as indicated by the grey
           bar. The $\psi$--dependence of the SOC--TW--transition $r^*$
           (solid line, filled squares), the TW saddle node bifurcation
           $r_{\rm TW}^s$ (dotted line, filled circles), and the
           saddle node bifurcation of the SOC branch $r_{\rm SOC}^s$
           (dashed line, open squares) is shown in a double logarithmical
           plot. Parameters are $L=0.01$ and $\sigma=10$.}
\label{fig:5}
\end{figure}
\end{document}